\documentstyle[aas2pp4]{article}
\lefthead{Kishimoto}
\righthead{The nuclear location and 3D view of NGC 1068}

\newcommand{\dgr}{^{\circ}}
\newcommand{\hst}{{\it HST}}
\newcommand{\chisq}{$\chi^{2}$\ }

\newcommand{\cmmax}{${\rm [CM]}_{\rm max}$}
\newcommand{\snmin}{${\rm [S/N]}_{\rm min}$}
\newcommand{\?}{\phantom{-}}
\newcommand{\s}{\phantom{0}}
\newcommand{\rp}{r_{p}}
\newcommand{\rc}{R}
\newcommand{\capsize}{\small}

\begin{document}

\title{The Location of the Nucleus of NGC 1068 and the 
Three-dimensional Structure of Its Nuclear Region \footnotemark[1]}

\footnotetext[1]{Based on observations made with the NASA/ESA Hubble 
Space Telescope, obtained from the data Archive at the Space Telescope 
Science Institute, which is operated by the Association of 
Universities for Research in Astronomy, Inc., under NASA contract NAS 
5-26555.}

\author{Makoto Kishimoto}
\affil{Department of Astronomy, Faculty of Science, Kyoto University\\
           Sakyo-ku, Kyoto 606-8502, Japan}

\bigskip

\begin{center}
{\it To appear in The Astrophysical Journal, June 1999}
\end{center}

\begin{abstract}
The \hst\ archival UV imaging polarimetry data of NGC 1068 is 
re-examined.  Through an extensive estimation of the observational 
errors, we discuss whether the distribution of the position angles 
(PAs) of polarization is simply centrosymmetric or not.  Taking into 
account the effect of a bad focus at the time of the observation, we 
conclude that, within the accuracy of \hst/FOC polarimetry, the PA 
distribution is completely centrosymmetric.  This means that the UV 
polarization originates only from scattering of the radiation from a 
central point-like source.

However, our analysis shows that the most probable location of the 
nucleus is only $\sim 0.''08$ ($\sim$ 6 pc) south from the brightest 
cloud called ``cloud~B''.  The error circle of 99\% confidence level 
extends to cloud~B and to ``cloud~A'' which is about $0.''2$ south of 
cloud~B. By this FOC observation, Cloud~B is only marginally rejected 
as the nucleus.

Assuming that the UV flux is dominated by electron-scattered light, we 
have also derived a three-dimensional structure of the nuclear region.  
The inferred distribution suggests a linear structure which could be 
related to the radio jet.

\end{abstract}


\section{Introduction}\label{s-intro}

The nucleus of the Seyfert 2 galaxy NGC 1068 is now firmly believed to 
be obscured from direct view.  This was clearly inferred by Antonucci 
\& Miller (1985) from a spectropolarimetric study, and now one of the 
major problems in this galaxy is to locate this hidden nucleus very 
accurately on the high-resolution images taken by the \hst\ and on 
VLBI radio maps.  This is particularly important, since the 
investigation of the physical conditions and kinematics of the nuclear 
vicinity is greatly influenced by the exact location of the hidden 
nucleus.

Several authors have addressed this problem, but most of the nuclear 
positions determined are indirect ones in the sense that they are from 
lower resolution images than the \hst\ images or VLBI maps, and these 
positions are slightly different from one another (see e.g. Thatte et 
al.  1997 and references therein).  The only exception so far for 
pinpointing the location directly on the \hst\ high-resolution images 
is to use imaging polarimetry data.  The nucleus can be determined as 
the center of the centrosymmetric distribution of the position angle 
(PA) of polarization, which is expected to be observed if the 
radiation from the nucleus is being scattered by the surrounding gas.  
(We refer to this case as a `point-source scattering case' hereafter.)  
Capetti et al.  (1995a, b) have determined the nuclear location by 
this method.  However, in their PA map, clear deviations from the 
centrosymmetric pattern are seen, and they did not state whether they 
are real, or discuss the observational error in PA at each position of 
the image.  If the deviations are real, the nuclear position 
determined by them would not be very accurate and reliable.

In this paper, we extend their work to discuss these deviations 
through the extensive estimation of the observational errors, and then 
to re-determine the location of the nucleus more accurately within a 
convincing error circle.  Our result is that the most probable 
location of the nucleus has moved to the north by $\sim 0.''2$, 
significantly larger than the quoted error of Capetti et al.  (1995b).

The second objective of this paper is to derive the three-dimensional 
structure of the nuclear region.  The \hst\ images show that this 
region has a knotty and filamentary structure.  If the UV radiation 
from each knot is dominated by the scattered radiation, the 
polarization degrees of the clouds provide the scattering angle at 
each cloud.  Hence, we can derive the three-dimensional distribution 
of these clouds with respect to the nucleus.

We describe the data in \S~\ref{s-data} and error estimation in 
\S~\ref{s-data-err}.  In \S~\ref{s-exam}, we examine the PA 
distribution, and in \S~\ref{s-exam-nuc} we discuss the location of 
the nucleus.  Then in \S~\ref{s-3D}, we derive the three-dimensional 
distribution of the scatterers.  We discuss these results in 
\S~\ref{s-disc} and our conclusions are presented in \S~\ref{s-conc}.  
In the appendix, we summarize the method of error estimation.  We 
assume a distance of 14.4 Mpc to NGC~1068 in this paper (Tully 1988), 
corresponding to a scale of $1'' \simeq 70$ pc.

\begin{deluxetable}{lllr@{\quad\quad}rl}
	\footnotesize
	\tablecaption{\hst/FOC archival data used\label{table-data}}
	\tablewidth{0pt}
	\tablehead{
	\colhead{Rootname}    & \colhead{Filter 1}    & \colhead{Filter 2} & 
	\colhead{Exp.time (sec)}  & \colhead{Date of obs.} & 
	\colhead{Description}
	}
	\startdata
	x274020at  &  F253M  &  POL0    &  1796.625  &  Feb 28, 1995 & UV continuum\\
	x274020bt  &  F253M  &  POL0    &   341.625  &  Feb 28, 1995 &\\
	x274020ct  &  F253M  &  POL0    &  1451.625  &  Feb 28, 1995 &\\
	x274020dt  &  F253M  &  POL60   &   748.625  &  Feb 28, 1995 &\\
	x274020et  &  F253M  &  POL60   &  1044.625  &  Feb 28, 1995 &\\
	x274020ft  &  F253M  &  POL60   &  1201.625  &  Feb 28, 1995 &\\
	x274020gt  &  F253M  &  POL60   &   591.625  &  Feb 28, 1995 &\\
	x274020ht  &  F253M  &  POL120  &  1608.625  &  Feb 28, 1995 &\\
	x274020it  &  F253M  &  POL120  &  1796.625  &  Feb 28, 1995 &\\
	x24e0102t  &  F501N  &  \nodata &  1196.000  &  Jan 10, 1994 & [OIII]\\
	x24e0103t  &  F501N  &  F4ND    &  1196.000  &  Jan 10, 1994 &\\
	x24e0106t  &  F253M  &  \nodata &  1196.000  &  Jan 10, 1994 & UV continuum\\
	\enddata
\end{deluxetable}

\section{The archival data}\label{s-data}
\subsection{Reduction procedure}\label{s-data-red}
The archival \hst\ data used are summarized in Table~\ref{table-data}.  
The polarimetry data were obtained on February 28, 1995.  These data 
have been published by Capetti et al.(1995b).  The central small 
portion was in the 10\%-level nonlinear regime, so a
flat-field linearity correction was applied, but we confirmed that 
this correction has no significant effect on our analysis, by 
implementing the same analysis below on the data without a linearity 
correction.  The data were subsequently processed in the standard 
manner to correct for geometric distortion and flat-field response.  
The reseau marks were removed using neighboring pixels.  The 
background subtraction was implemented using the outermost regions of 
the images.

These data were obtained after the COSTAR deployment, but at the time 
of this polarimetry observation \hst\ had an extraordinarily poor 
focus due to a large movement of the secondary mirror, the effect of 
which will be discussed in detail below.  Therefore, the images 
through different polarizers (POL0, POL60, POL120), which are known to 
be slightly shifted relative to one another, were registered by using 
the image shift values of calibrations by Hodge (1993, 1995), not by 
using point-like sources in the outer region.  Then the images were 
scaled according to the exposure times and the transmittances of the 
polarizers and the F253M filter.  (The filter transmits the UV 
radiation around 2400-2700\AA.) Finally these images were combined to 
obtain the Stokes parameter $I$, $Q$, $U$ images.

Using several archival \hst/FOS spectra taken with $0.''3$ aperture, 
we have estimated the amount of the emission-line contamination in the 
F253M image to be around the $10 \sim 15\%$ level.  This includes, 
however, the broad FeII lines which are also a part of the scattered 
light in addition to the scattered continuum (Antonucci, Hurt, \& 
Miller 1994).  Therefore, the effect of the line contamination on the 
polarization should be much smaller.

We have also used the archival images of NGC~1068 taken through the 
F501N filter, i.e., the [OIII] image, and through the F253M filter but 
without polarizers, the focus of which seem to be fine.  They were 
obtained on January 10, 1994, also after the COSTAR deployment.  They 
have been published by Macchetto et al.(1994).  The images were 
processed through the same procedure as above, but in the F501N image, 
the central nonlinear and saturated portion was filled with the 
appropriately scaled image taken with the F4ND filter.  The pattern 
noise associated with the FOC nonlinearity was removed by Fourier 
filtering.  The registration of these two images was carried out using 
point-like sources in the outer region, with the uncertainty of less 
than 1.4 pixel.  The registration of this F253M image and the $I$ 
(total intensity) image from the polarimetry data was done by taking a
cross-correlation, because of the bad focus in the $I$ image.  We also 
tried this registration using point sources and found the results 
coincide within 1 pixel.  

Both of the observations were implemented in normal $512 \times 512$ 
mode, where the pixel size is $0.''014 \times 0.''014$ and the 
field of view is $7'' \times 7''$.

\subsection{\hst\ Focus and Degradation of the Image}
\label{s-data-focus}
Due to the bad focus in the imaging polarimetry observation, the 
images have a significant blurring.  The effect on the polarization 
analysis should be significant especially at the region where the 
gradient of the polarized flux distribution is large.  We first 
estimate the extent of this blurring by using the fine-focus 
(non-polarimetric) F253M image.

\begin{figure}[bt]
\epsscale{1.0}
\plotone{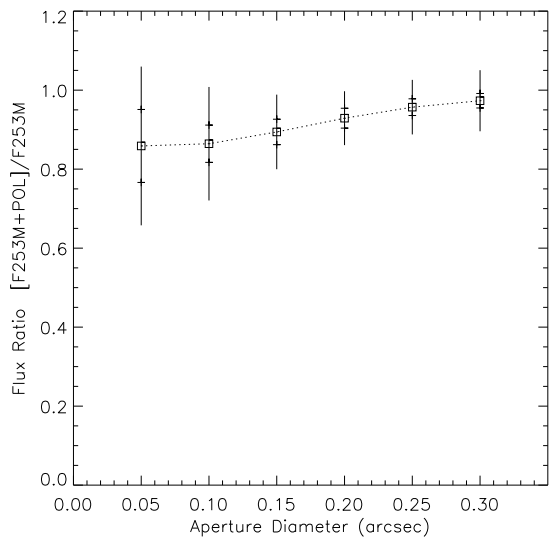} 
\figcaption{\label{fig-cm} {\capsize 
The averaged ratio of the $I$ image to the sharp UV image over various 
aperture diameter is shown as squares, connected by dotted lines.  
The vertical line on each square represents the standard deviation of 
the scatter from each averaged ratio. The expected amount of the 
error in the ratio due to statistical noise is shown in horizontal 
ticks.}}
\end{figure}

To avoid confusion, we call this F253M image without polarizers a 
``sharp'' UV image, and the $I$ image of the polarimetry simply an $I$ 
image.  Synthetic aperture photometry was carried out to take the 
ratio of the $I$ image to the sharp UV image at various aperture 
diameters, and the results on a few tens of local maxima over the 
image were averaged.  The results are shown in Figure~\ref{fig-cm}.  
The scatter of the ratio for each aperture is shown as a vertical 
line.  The expected amount of the error in the ratio due to 
statistical noise is also shown with horizontal ticks.  As we expect 
that the local peak intensity is lowered by the blurring, the average 
ratio is certainly less than unity and it becomes smaller with smaller 
aperture, about 0.9 at $0.''15$ aperture, down to $\sim$0.85 at 
$0.''05$ (though the statistical significance of these average ratios 
becomes lower with smaller aperture).  The regions in the $I$ image 
with this ratio larger than $\sim 1.0$ have significant `leaks' from 
the neighboring bright regions.

To infer correctly the degradation of the point spread function (PSF) 
of this observation, we have estimated the amount of the secondary 
mirror motion in the following way.  We constructed model PSFs using 
TinyTim (Krist 1993) with a range of secondary mirror movement and 
convolved the sharp UV image with those PSFs.  Then we compared them 
to the $I$ image which is convolved with the PSF of nominal focus for 
equivalence.  We calculated the ratio of these two by the same 
synthetic aperture photometry as in Figure~\ref{fig-cm}.  It is 
expected that this ratio becomes unity for all aperture sizes when the 
appropriate amount of the secondary mirror shift is taken for the 
model PSF. We found that this was the case for a shift of 
about $9 \sim 10$ microns.  The model PSF at this focus indicates that 
the FWHM of the PSF was $0.''10 \sim 0.''15$ at the time of the 
observation.

Therefore, we decided to implement our analysis in three ways; 
(A) $10 \times 10$ pixel binning, corresponding to $\sim 0.''15 \times 
0.''15$, (B) $20 \times 20$ pixel binning, (C) $0.''15$ aperture 
diameter synthetic photometry.  The size for the cases (A) and (C) 
could be too small, but the effect of small aperture will be taken 
into account in the error estimation.  For case (C), the aperture 
centers are selected as the local peaks in the $I$ image, which are 
thought to represent the positions of the resolved clouds in this 
nuclear region.

\section{Error estimation}\label{s-data-err}
In order to examine the PA distribution accurately, we need extensive 
estimation of the observational error in the \hst/FOC polarimetry.  We 
describe major ideas here and summarize other details in appendix.

We estimate the error in polarization degree and PA by considering the 
following four sources of errors that are expected to be major among 
various error sources.

(1) Statistical error. Poisson noise is assumed.

(2) Uncertainty from the image registrations of three polarizer 
images.  As described in \S~\ref{s-data-red}, we have registered the 
images using the calibrated values of Hodge (1995) and their 
wavelength dependence by Hodge (1993).  The calibrated image shifts 
have an uncertainty of about $\pm 0.3$ pixel.  Therefore we shifted 
POL60 and POL120 images by this amount along the image x and y axes 
relative to POL0 image and calculated the resulting change of the 
polarization.

(3) Polarizer axes uncertainty.  The transmission axes of each 
polarizer have nominal values of $180\dgr$, $60\dgr$, $120\dgr$ with 
respect to the $+x$ axis of the image, and those values have about 
$\pm3\dgr$ uncertainties (Nota et al.  1996).  The errors in the 
calculated polarization due to this uncertainty were estimated by 
considering the dependence of the calculated polarization on the 
polarizer axes' angles (see appendix for more detail).

(4) We consider the uncertainties from the following error sources as the 
uncertainties in the ``correction factors'' to be multiplied to each image 
through each polarizer.

\begin{itemize}

  \item[(i)] Each polarizer, especially POL60, has different 
  wavelength dependence of transmittance, so when the spectrum of the 
  object changes from place to place, each polarizer's image has to be 
  multiplied by a different factor to correct for the different 
  effective throughput.  By assuming a power-law source spectrum and 
  varying the power-law index in a reasonable range, we estimate 
  that this has effects of less than about 1\% in the correction 
  factors.

  \item[(ii)] Differences in PSFs through each polarizer result in 
  uncertainties in these multiplying factors locally.  We estimate 
  this effect from the result given by Hodge (1995), who has 
  investigated the differences in the fluxes of unpolarized 
  point sources through different polarizers over various aperture 
  sizes.  Although this is for data with nominal focus, the maximum 
  uncertainties for our out-of-focus data can be estimated for each of 
  our binning size (A), (B) and (C) above, from the flux differences 
  of smaller aperture size in Hodge's result.  For cases (A) 
  and (C), the relative discrepancies between the fluxes through each 
  polarizer are expected to be less than 3 to 4 \%, reading from 
  Hodge's plots for 3 to 5 pixel aperture radius, and for case (B) 
  less than 2 to 3\%, from the same plots for $\sim 7$ to 10 pixel 
  aperture radius.  The fact that Hodge's result is for point 
  sources while our sources are more extended, gives us additional 
  support that the values above can be considered as the maximum 
  uncertainties.

  \item[(iii)] Flat-fielding uncertainty.  The flat fields used by the 
  FOC standard calibration are heavily smoothed and they do not 
  correct for the fine scale features in the original, unsmoothed flat 
  fields.  We have estimated the resulting flat-fielding uncertainties 
  by calculating the deviations of the unsmoothed flat field from the 
  smoothed one, for the used portion of the image.  The results are 
  about 3\% for the cases (A) and (C),
  2\% for the case (B).

\end{itemize}

Taking the sum of the squares of these (i)$\sim$(iii) factors, we 
consider uncertainties in each multiplying factor as 5\% for cases (A) 
and (C), 4\% for case (B), at most.  The effect of these uncertainties 
on the polarization were calculated from the dependence of the 
polarization on these multiplying factors (see appendix for more 
detail).

The total error was estimated by combining in quadrature all the 
uncertainties above from (1) to (4).  Source (4) is usually larger 
than (2) and (3).  Typically, the total estimated error in the 
polarization degree, $\sigma_{P}$, in the regions of sufficient 
statistical S/N (the portion of the data used in the analysis below) 
was about $4.5 \sim 5.5\%$ for the cases (A) and (C), while about $3.5 
\sim 4.5\%$ for the case (B).  If the statistical error is excluded, 
the quadratic sum of the other three was about $4.5\%$ for cases (A) 
and (C), about $3.5\%$ for case (B).

\section{Examination of Position Angle Distribution}\label{s-exam}

As shown in Capetti et al.  (1995b), the overall PA distribution is 
quite centrosymmetric, quite close to a `point-source scattering 
case'.  However, clear deviations from the centrosymmetric pattern are 
seen in some regions, especially around the very center (see their 
Fig.2).  We now examine these deviations and discuss whether the data 
are really consistent with this simple model, based on the PA errors 
estimated in the previous section and also on the investigation of the 
bad-focus effect discussed in \S~\ref{s-data-focus}.

\subsection{Construction of \chisq image}\label{s-exam-chisq}
We evaluate whether the data are really consistent with point-source 
scattering by calculating the \chisq value.  For each $i$th binned 
pixel or aperture, denote the PA data as $\theta_{\rm PA}(i)$, and the 
ideal centrosymmetric PA with a certain symmetric center 
$(x_{c},y_{c})$ as $\theta_{\rm ideal}(i\ ; x_{c},y_{c})$.  We write 
the \chisq value for $(x_{c},y_{c})$ as
\begin{equation}
	\chi^{2}(x_{c},y_{c}) 
	  = \sum_{i}^{\rm valid \thinspace data} 
	    \frac{ \{ \theta_{\rm PA}(i) - 
	    \theta_{\rm ideal}(i\ ; x_{c},y_{c}) \}^{2} 
	    }{\sigma_{\theta_{\rm PA}}(i)^{2}},
	\label{eq-chisq}
\end{equation}
using the total estimated error in PA, $\sigma_{\theta_{\rm PA}}$, 
described in the previous section.  The \chisq values for various 
symmetric centers are calculated and a `\chisq image' is constructed, 
in which the image value at a certain point represents the \chisq 
value with this point being the symmetric center. 

\begin{figure}[bt]
\epsscale{1.0}
\plotone{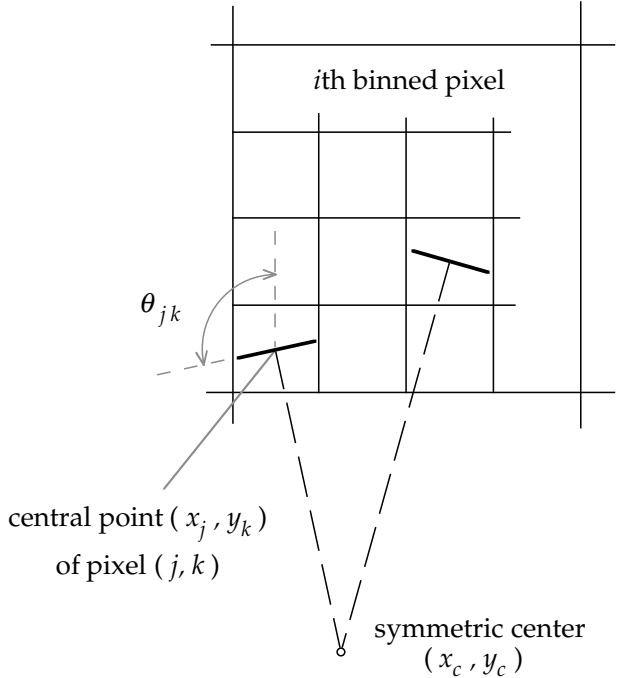}
\figcaption{\label{fig-idealPA} {\capsize
Illustrates how the centrosymmetric PA for 
each binned pixel is calculated using the polarized flux distribution 
within the binned pixel as weight.  Thick solid vectors represent the 
PAs perpendicular to the direction to the symmetric center.}}
\end{figure}

The ideal centrosymmetric PA for one large binned pixel or aperture 
depends on, in addition to the direction to the symmetric center, the 
polarized flux distribution within that pixel.  Therefore we used the 
original, not binned, distribution of polarized flux within the binned 
pixel or aperture for the calculation of the \chisq value.  That is, 
we calculate the ``polarized-flux-weighted'' ideal centrosymmetric PA 
(actually, weighted further by the reciprocal of statistical error of 
polarized flux, to avoid having too much weight on the pixels with low 
S/N polarized flux) as
\begin{equation}
	\theta_{\rm ideal}(i\ ; x_{c},y_{c}) = 
	  \frac{1}{2}\arctan \frac{U_{\rm ideal}(i)}{Q_{\rm ideal}(i)},
	\label{eq-idealPA}
\end{equation}
where
\begin{eqnarray}
	Q_{\rm ideal}(i) 
	  & = & \frac{\sum_{j,k}^{\rm {\it i}th\ bin}
	  w_{jk} I_{P}(j,k) \cos2\theta_{jk}}{
	  \sum_{j,k}^{\rm {\it i}th\ bin} w_{jk}}, \\
	U_{\rm ideal}(i) 
	  & = & \frac{\sum_{j,k}^{\rm {\it i}th\ bin}
	  w_{jk} I_{P}(j,k) \sin2\theta_{jk}}{
	  \sum_{j,k}^{\rm {\it i}th\ bin} w_{jk}}, \\
	w_{jk} 
	  & = & \frac{1}{\sigma_{I_{P}}^{\rm stat}(j,k)}, \\
	\theta_{jk}
	  & = & \arctan\frac{y_{k}-y_{c}}{x_{j}-x_{c}}.
\end{eqnarray}
The quantities $I_{P}(j,k), \sigma_{I_{P}}^{\rm stat}(j,k), 
(x_{j},y_{k})$ are the polarized flux, statistical error in the 
polarized flux, and central point, for the original pixel $(j,k)$ 
within the $i{\rm th}$ binned pixel, respectively.  These are 
illustrated in Figure~\ref{fig-idealPA}.  The summations are taken 
over all pixels $(j,k)$ within the $i{\rm th}$ binned pixel.  The 
angle $\theta_{jk}$ corresponds to the ideal centrosymmetric PA for 
the center of the pixel $(j,k)$.

This correction in the ideal PA (the difference between the 
polarized-flux-weighted ideal PA and the centrosymmetric PA which is 
calculated simply for the center of the binned pixel or aperture) was 
found to be very important in the central bright and knotty region.  
In this region, the amount of the correction was larger than the 
total error of PA estimated in \S~\ref{s-data-err}.

\subsection{The effect of bad focus on the evaluation of \chisq}
\label{s-exam-mask} 

The extraordinarily degraded PSF of these polarimetry data can affect 
the PA distribution significantly, especially in the pixels 
surrounding the bright clouds, even with relatively large pixel 
binning.  We eliminate this effect from the analysis by masking out 
the binned pixels or apertures which are suspected to have significant 
`leak' from the surrounding pixels.  If we take the ratio of the $I$ 
image to the sharp UV image for each bin or aperture, the pixels or 
apertures which are not affected by the surrounding regions should 
have this ratio less than around unity, which can also be down to 
$\sim 0.9$ for $0.''15$ aperture as shown in Figure~\ref{fig-cm} and 
discussed in \S~\ref{s-data-focus}.  We call this ratio here a 
``contamination measure''.  We should mask out the region where this 
contamination measure is much larger than unity, so we set an 
appropriate threshold on this contamination measure, which we denote 
as \cmmax.  In addition, we use only the region where the polarization 
has been detected with high statistical S/N. We set an appropriate 
threshold on $P/\sigma_{P}^{\rm stat}$ (statistical S/N of $P$), which 
we write as \snmin.

\begin{figure}[bt]
\epsscale{1.0}
\plotone{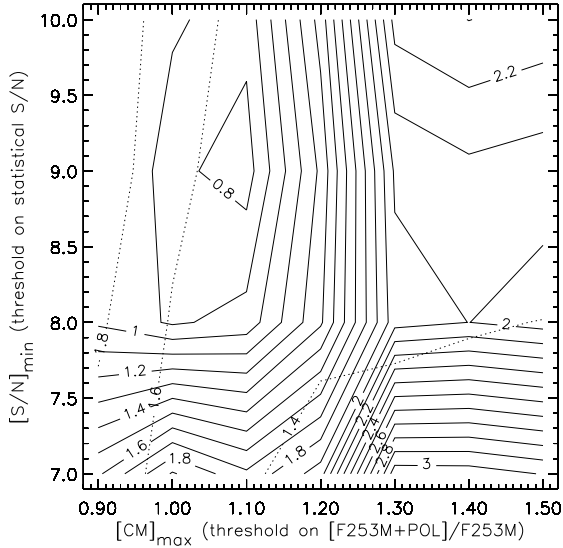}
\figcaption{\label{fig-csm} {\capsize
The map of the minimum reduced \chisq for 
case (A) is drawn in solid contours on the plane of the thresholds on the 
contamination measure (ratio of the $I$ image to the sharp UV 
image) and statistical S/N ($P/\sigma_{P}$).  Also shown in dotted 
contours are the cut-off values of the reduced \chisq with 99\%
confidence level for
each degree of freedom in each set of thresholds.}}
\end{figure}

The result is that, choosing appropriate thresholds on these two 
factors, we have obtained minimum reduced \chisq (the \chisq divided 
by degrees of freedom) of slightly less than unity for all three 
binning cases from (A) to (C).  This indeed means that these 
polarimetry data are totally consistent with point-source 
scattering, within the accuracy of the FOC polarimetry.  In 
Figure~\ref{fig-csm}, we show the map of the minimum reduced \chisq 
value in the case (A), for various sets of \cmmax\ and \snmin.  The 
\chisq becomes smaller for larger \snmin\ and smaller \cmmax, and we 
clearly see the \chisq becomes minimum and almost constant for \snmin\ 
$\geq 8$ and \cmmax\ $\leq 1.1$, with its value indicating the fit is 
very good.  The dotted-line contours indicate the cut-off value of the 
reduced \chisq with 99\% confidence level for each degree of freedom 
in each set of \snmin\ and \cmmax.  The \chisq also becomes stable for 
\snmin\ $\geq 8$ and \cmmax\ $\geq 1.3$, but the reduced \chisq value 
for this region is rather hard to accept, since it is much larger than 
the cut-off value.  Furthermore, there is a clear transition between 
these two regions where the gradient of \chisq is large and \chisq 
switches to an acceptable value when \cmmax\ becomes smaller.  Similar 
results have been obtained for the cases (B) and (C).  Therefore we 
conclude that the PA distribution is consistent with a point-source 
scattering only when the contaminated regions are excluded.

\begin{figure}[bt]
\epsscale{1.0}
\plotone{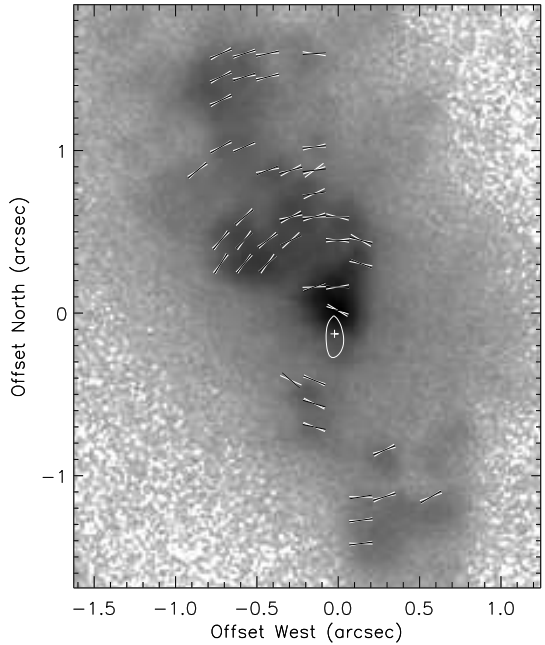}

\figcaption{\label{fig-PA-A} {\capsize The PA distribution with $10 
\times 10$ pixel binning [case (A)] is shown on the $I$ image of the 
UV polarimetry.  The position of minimum \chisq is shown as a plus 
sign, and the error circle of 99\% confidence level is also drawn.  
The two white vectors for each binned pixel indicate $\theta_{\rm PA} 
\pm \sigma_{\theta_{\rm PA}}$ and the black vector shows 
polarized-flux-weighted ideal centrosymmetric PA with the minimum 
\chisq position being the symmetric center.  The image is in log 
scale, and has been cut at 0.01\% of the peak intensity.}}

\end{figure}

\begin{figure}[bt]
\epsscale{1.0}
\plotone{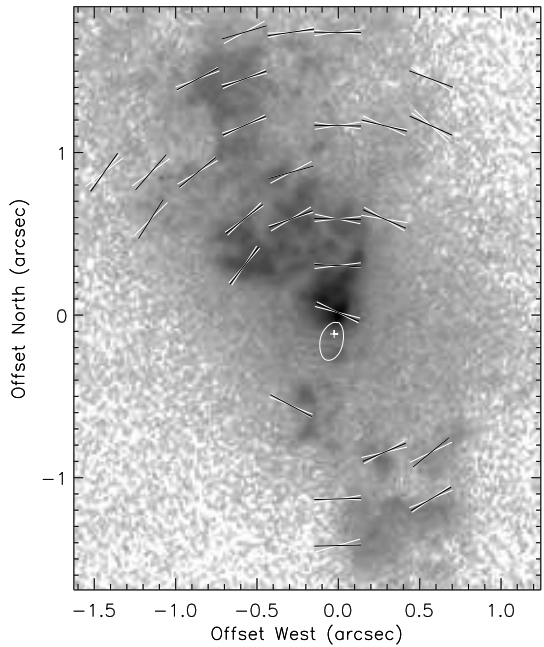}
\figcaption{\label{fig-PA-B} {\capsize
The same as Fig.\ref{fig-PA-A}, but with 
$20 \time 20$ pixel binning [case (B)], and the underlying image is 
the ``sharp UV image'', which was taken when the \hst\ focus was fine.  
The image is in log scale, and has been cut at 0.01\% of the peak
intensity.}}
\end{figure}

\begin{figure}[bt]
\epsscale{1.0}
\plotone{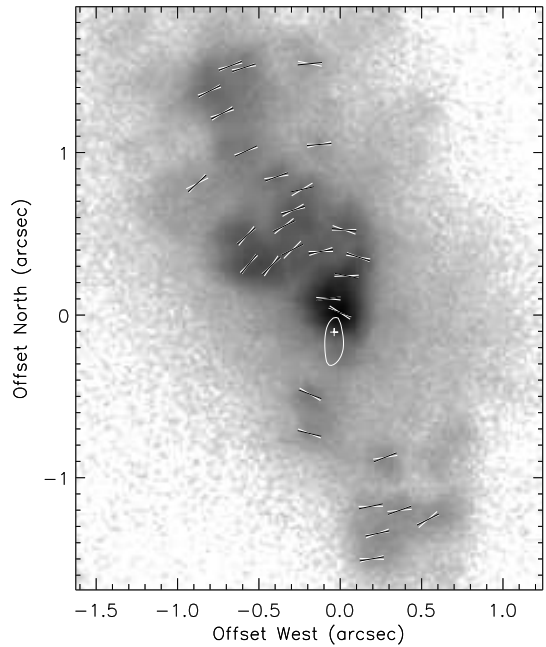}
\figcaption{\label{fig-PA-C} {\capsize
The same as Fig.\ref{fig-PA-A}, but for 
the case of aperture photometry on the local peaks with $0.''15$ 
diameter [case (C)].  The underlying image is the $I$ image from the 
imaging polarimetry.  It is in log scale, and has been cut at 0.1\% of 
the
peak intensity to show the bright regions more clearly.}}
\end{figure}

\section{The location of the nucleus}
\label{s-exam-nuc}
Now that the point-source scattering fit is proved to be 
acceptable, we can discuss the most probable location of the nucleus 
as the position of the minimum \chisq point, and its error as the 
contour of the \chisq image.

\subsection{Minimum \chisq point}

We show the PA distribution and the point with the \chisq being 
minimum for each case (A), (B), and (C) in Figure~\ref{fig-PA-A}, 
\ref{fig-PA-B}, and \ref{fig-PA-C}, respectively.  The observed PAs 
for each binned pixel or aperture are shown in two white vectors that 
correspond to $\theta_{\rm PA} \pm \sigma_{\theta_{\rm PA}}$, where 
$\sigma_{\theta_{\rm PA}}$ is the total estimated error in PA. Also 
shown in black vectors are the polarized-flux-weighted ideal 
centrosymmetric PA with the symmetric center being the minimum \chisq 
point, marked as a plus, for each binning case.  For these figures, we 
have chosen \cmmax\ to be 1 in all cases, which is considered to be a 
natural constraint on the contamination measure (though we can also 
take the values of up to $\sim 1.1$ to allow for the statistical 
noise, as seen in Figure~\ref{fig-csm}).  We have taken the lowest 
\snmin\ in each binning case that could yield acceptable \chisq 
values, in order to include as large an area as possible.  The minimum 
reduced \chisq value was found to be 0.88 for 41 degrees of freedom 
with \snmin\ = 8 in case (A), 0.80 for 26 degrees with \snmin\ = 7 in 
case (B), and 0.74 for 28 degrees in case (C).  The underlying image 
in Figure~\ref{fig-PA-A} and Figure~\ref{fig-PA-C} is the $I$ image, 
while the sharp UV image is presented in Figure~\ref{fig-PA-B} for 
comparison.

\begin{figure}[bt]
\epsscale{1.0}
\plotone{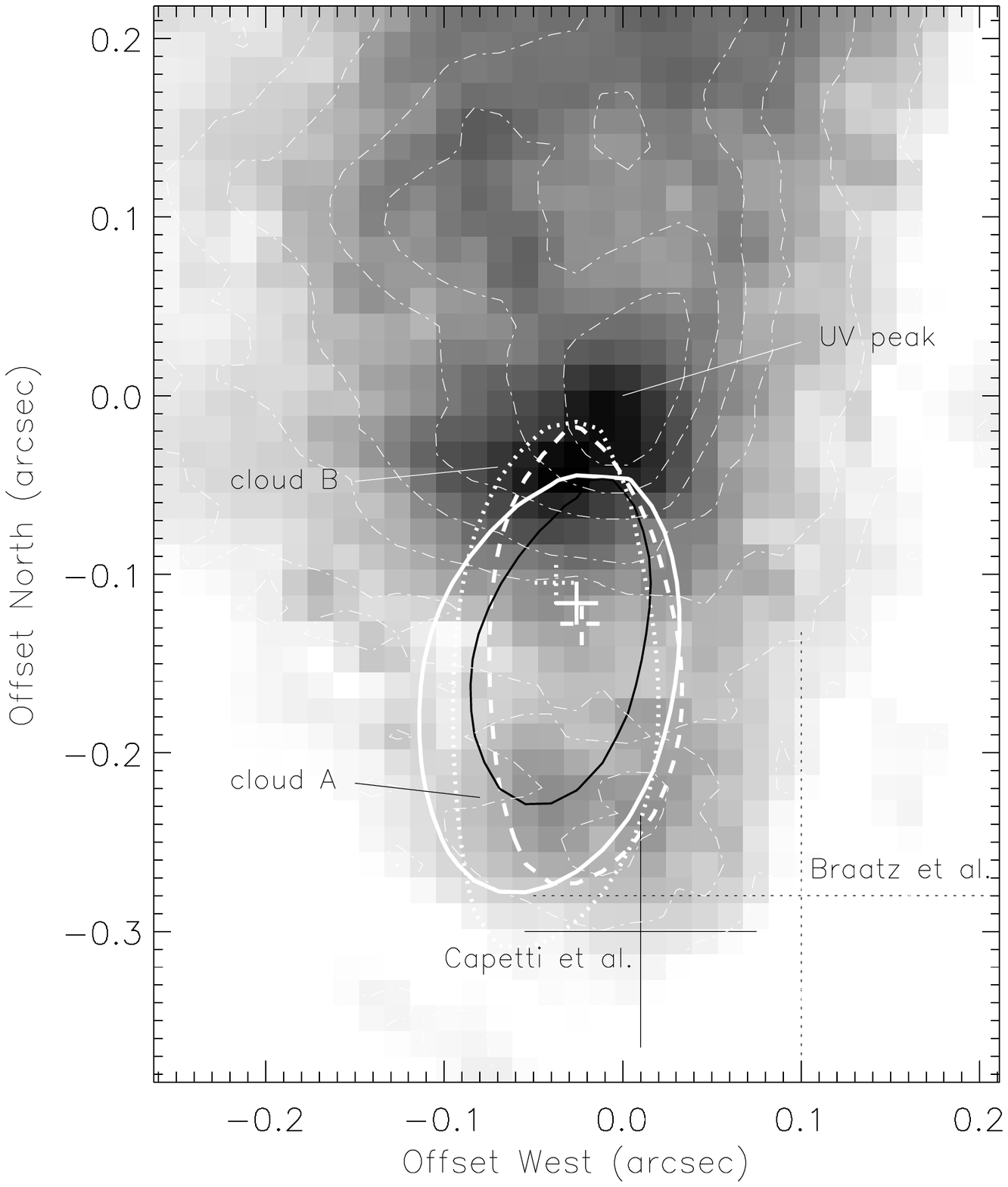}

\figcaption{\label{fig-nuc} {\capsize The enlarged view of the central 
$\sim 0.''5$ region.  The positions for minimum \chisq points and 99\% 
confidence level contours are shown in white dashed, solid, and dotted 
line for cases (A), (B), and (C), respectively.  The black solid 
contour represents 99\% confidence level error circle for case (B) 
with smaller uncertainty in the ``correction factor''.  Underlying is 
the F501N [OIII] image, and the dash-dot contours represent the sharp 
UV image.  Both are in log scale.  The thin solid and dotted plus sign 
indicates the nucleus location suggested by Capetti et al.  (1995b) 
and Braatz et al.  (1993), respectively, with their sizes being the 
errors estimated by the respective authors.}}

\end{figure}

Compared to the previous result from the same data (Capetti et al.  
1995b), the minimum \chisq point, i.e., the most probable location of 
the nucleus, is very close to the UV brightest cloud, in all cases 
from (A) to (C), as seen in Figure~\ref{fig-nuc} covering just the 
central $\sim 0.''5$ region.  The white dash-dot contour represents 
the UV image, while underlying is the [OIII] image through the F501N 
filter.  Our minimum \chisq points shown in small plus signs are only 
about $0.''12$ south from the UV peak at the origin of the 
coordinates, whereas Capetti et al.  (1995b) have located the nucleus 
at $0.''3$ south from the UV peak (the value is actually taken from 
Capetti, Macchetto, \& Lattanzi 1997), shown as a large plus sign with 
its size being their estimated error.

The reason for this difference is quite clear.  Corresponding to the 
``transition'' of the minimum \chisq value in Figure~\ref{fig-csm} 
from \cmmax\ of 1.3 to 1.1, we have found that the position of the 
minimum \chisq point experiences the transition from $\sim (0.''01, 
-0.''23)$, which is almost within the error of the Capetti et al.  
point, to our points $\sim (-0.''03, -0.''12)$.  We found two regions 
on the plane of (\cmmax, \snmin) where the position of the minimum 
\chisq point is stable, just corresponding to the two regions in 
Figure~\ref{fig-csm} where the minimum \chisq values are almost 
constant.  These results are for case (A), but we see quite similar 
results for cases (B) and (C).  Therefore, the difference between the 
two locations is expected to arise from the masking-out procedure for 
the regions contaminated by the focus effect.

\begin{figure}[bt]
\epsscale{1.0}
\plotone{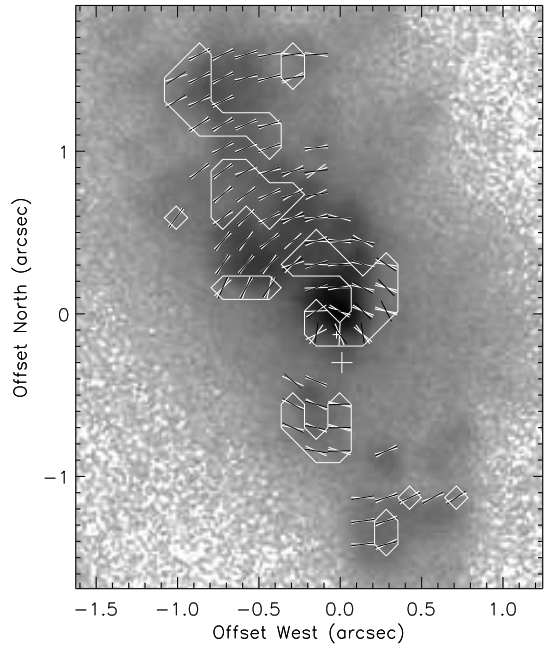}
\figcaption{\label{fig-mask} {\capsize
The same as Fig.~\ref{fig-PA-A} but the 
observed PAs in the masked out region, enclosed by white lines, are 
also drawn.  The small plus sign is the same as in 
Fig.~\ref{fig-PA-A}, while the large plus sign represents the location of the 
nucleus suggested by Capetti et al.  (1995b), with its size being their 
estimated error.}}
\end{figure}

The reason for the shift of this direction is also fairly clear.  We 
show the PA distribution in the masked-out region in 
Figure~\ref{fig-mask}, enclosed by white lines, in addition to the PA 
distribution already shown in Figure~\ref{fig-PA-A}.  The masked-out 
region, especially around the UV peak, seems to direct the nucleus to 
the south. So the minimum \chisq point moves to the north when 
these regions are excluded.

These PAs in the masked-out regions clearly cannot be fit by 
point-source scattering, and we have certainly shown that the 
minimum \chisq becomes acceptable only if we mask out these regions.  
The tendency is that the deviations in these regions can be 
explained by the contamination from the neighboring bright regions.  
Since our masking-out procedure is quite natural and reasonable, and 
this tendency gives us further support for our procedure, we 
conclude that the Capetti et al. point should be revised to a more 
northern point, indicated by our minimum \chisq positions.

\subsection{Error circle}
The error of the location of the symmetric center can be estimated as 
contours of the \chisq image.  The contour of 99\% confidence level 
has been drawn for each
binning method in Figures~$\ref{fig-PA-A}\sim\ref{fig-PA-C}$, and these 
three different contours are all shown with an enlarged scale in 
Figure~\ref{fig-nuc} in three different white circles.  These three 
are slightly different from one another.  This is partly due 
to the uncertainties in our estimation of PA errors, but the error 
circle for case (B) is smaller in the north-south direction partly 
due to the fact that the valid data points in case (B) are more 
extended to the east and west (Fig.~\ref{fig-PA-B}), while in 
cases (A) and (C) they are only extended to the north and south.

\begin{deluxetable}{lllll}
	\footnotesize
	\tablecaption{Positions of the Nucleus and Other Sources
	  \label{table-pos}}
	\tablewidth{0pt}
	\tablehead{
	\colhead{source name}   & \colhead{offset west $('')$\tablenotemark{a}} & 
	\colhead{offset north $('')$\tablenotemark{a}} & 
	\colhead{R.A. (J2000)} & \colhead{Dec. (J2000)} 
	}
	\startdata
	UV and optical peak & $\? 0.00$ & $\? 0.00$ & 
	$02^{\rm h} 42^{\rm m} 40.^{\rm s}711 \pm 0.^{\rm s}005$ & $-00\dgr 00' 47.''81 \pm 0.''08$
	\tablenotemark{b}\\
	hidden nucleus    & $-0.03 \pm 0.05$ & $-0.12 ^{+ 0.07}_{- 0.12}$ & 
	$02^{\rm h} 42^{\rm m} 40.^{\rm s}713 \pm 0.^{\rm s}006$ & $-00\dgr 00' 47.''93 ^{-0.''11}_{+0.''14}$
	\tablenotemark{c}\\
	S1 source         & $\? 0.02 \pm 0.10$ & $-0.13 \pm 0.10$\tablenotemark{b}& 
	$02^{\rm h} 42^{\rm m} 40.^{\rm s}710 \pm 0.^{\rm s}001$ & $-00\dgr 00' 47.''94 \pm 0.''02$
	\tablenotemark{d}\\
	\enddata
	\tablenotetext{a}{Offset from UV and optical peak.} 
	\tablenotetext{b}{From Capetti et al. 1997.}
	\tablenotetext{c}{Quadratic sums were taken to obtain errors.}
	\tablenotetext{d}{From Muxlow et al. 1996.}
\end{deluxetable}

As we described in \S~\ref{s-data-err}, the uncertainty in the 
``correction factor'' should be considered as the maximum possible, 
which might be indicated also by the fact that our minimum \chisq 
value is slightly smaller than unity.  We have also implemented the 
calculation of \chisq with a slightly smaller value for this 
uncertainty, and found that the minimum \chisq points and northern 
part of the error circles were almost stable, whereas the southern 
part of the error circles moved slightly to the north.  This shift is 
about $0.''05$ if we take the uncertainty in the correction factor to 
be 4\% for cases (A) and (C), and 3\% for case (B), instead of 5\% for 
(A) and (C), and 4\% for (B).  In Figure~\ref{fig-nuc}, the error 
circle of 99\% confidence level has been drawn as a black contour for 
the case (B) with 3\% uncertainty in the correction factor.  The 
minimum \chisq values were 1.16, 1.06, and 0.99 for the cases (A), (B), 
and (C), respectively, which are still low enough to accept the fit.  
Therefore, the error circle would be smaller in the southern part.

The brightest [OIII] cloud in Figure~\ref{fig-nuc} is called 
``cloud~B'' and the fainter cloud $0.''2$ to the south is called 
``cloud~A'' (Evans et al.  1991; see also Bland-Hawthorne et al.  
1997).  Cloud~B is slightly but significantly displaced from the 
brightest UV cloud, where the offset is about $0.''05$ at PA $\sim 
-37\dgr$.  These are consistent with the values noted by Macchetto et 
al.  (1994).  Note that the accuracy of the registration of these two 
images is estimated to be better than 1.4 pixel, corresponding to 
$0.''020$, as described in \S~\ref{s-data-red}.  The error circles 
suggest that the nucleus is located between clouds~A and~B. The 
projected distance between cloud~B and our minimum \chisq point is 
only $\sim 0.''08$ ($\sim$~6~pc).  Cloud~B is only marginally 
rejected as the symmetric center.  This means both that the quality of 
these FOC polarimetry data can only limit the location with this 
amount of error, and that the nucleus could be located just beside 
cloud~B.

For comparison, the location of the nucleus determined by Braatz et 
al.  (1993) as the peak position at $12.4 \mu{\rm m}$ has been drawn 
in Figure~\ref{fig-nuc}, with the size being their estimated error.  
This location is given with respect to the optical continuum peak, but 
we have used their value with respect to the UV peak instead, since 
the UV peak and optical peak are coincident within the accuracy of 
the registration of \hst/WFPC2 and FOC images (Capetti et al.  1997; 
Kishimoto 1998).  Our minimum \chisq positions are marginally within 
their error box.  Thatte et al.  (1997) have also located the nucleus 
as the peak of near-infrared emission, which is slightly north of, but 
almost identical to the position of Braatz et al.

In conclusion, allowing for the displacement of the minimum \chisq 
points and error circles of the four cases in Figure~\ref{fig-nuc}, we 
locate the nucleus $0.''12$ south and $0.''03$ east from the UV peak, 
with the error circle extending from this point about $0.''07$ to the 
north, $0.''12$ to the south, and $0.''05$ to the east and west.  In 
Table~\ref{table-pos}, we summarize this result with equatorial 
coordinates, using the result of absolute astrometry by Capetti et al.  
(1997) on the \hst\ continuum peak.

\section{Three-dimensional structure}\label{s-3D}
We have shown that the UV polarization is really consistent with 
point-source scattering and determined the location of the nucleus 
accurately. Based on these two results,  we infer the 
three-dimensional structure of the nuclear region, by using the 
polarization degrees as the indicator of scattering angles or 
`viewing angles' at each resolved cloud in the image.

\subsection{Assumptions}\label{s-3D-assume}
\begin{figure}[bt]
\epsscale{1.0}
\plotone{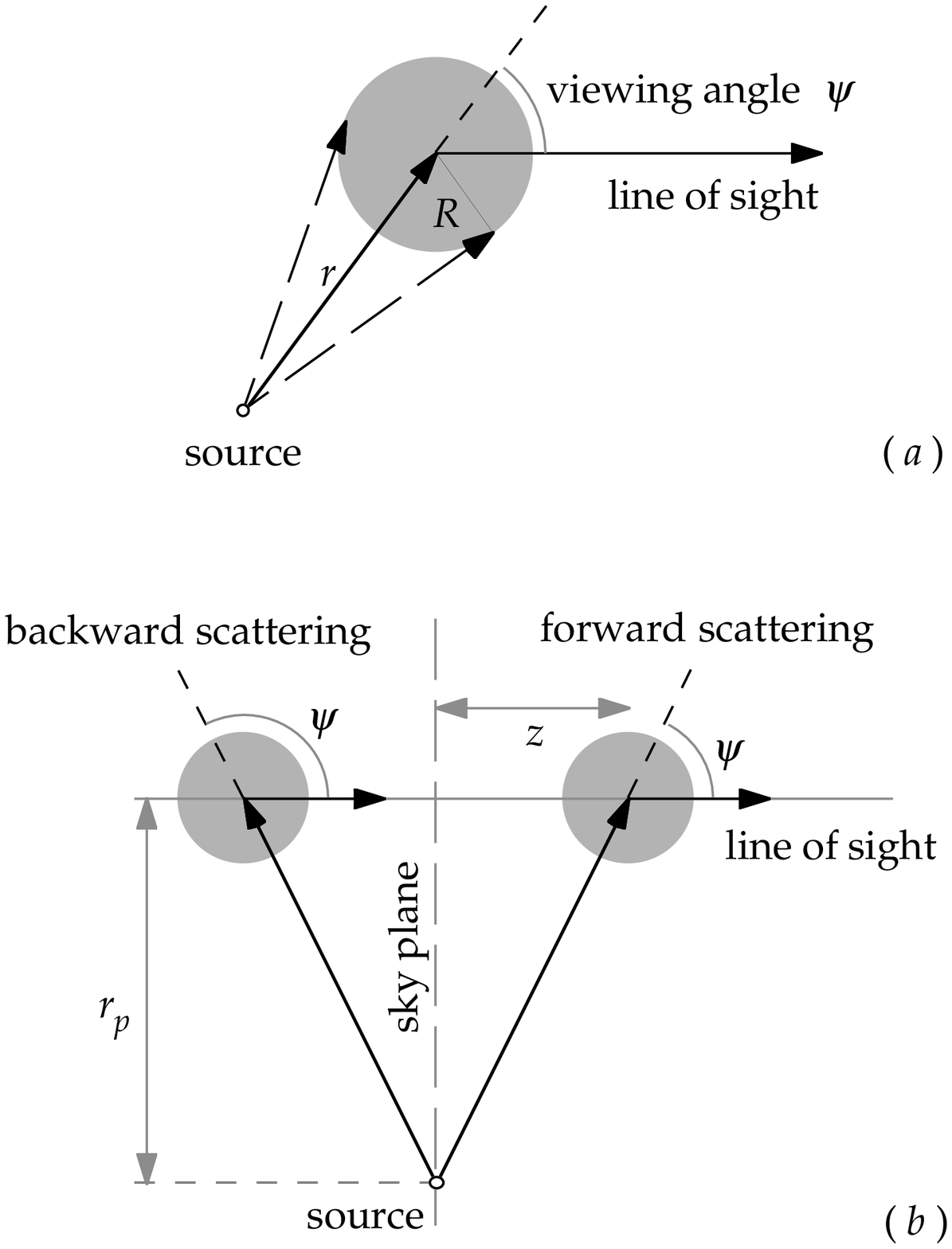}
\figcaption{\label{fig-clump} {\capsize
Illustrates the configuration of the 
cloud model.  (a) Definition of viewing angle of a spherical cloud.  
(b) The two `critical' positions determined from the polarization of 
the cloud.  }}
\end{figure}

Interpretation of the polarization degree is usually difficult, due to 
the existence of unpolarized, ``diluting'' radiation in terms of 
polarization, such as starlight.  In the nuclear region of NGC~1068, 
however, there is some evidence that suggests the diluting radiation 
in the UV range is fairly small, as described below.  If this is 
correct, we are in the unusual situation that we can infer the 
distribution of the nuclear resolved clouds three-dimensionally.  The 
polarization degree reflects the angle between our line of sight to 
the scatterers and the line of sight of those scatterers themselves to 
the illuminating source, which we call here `viewing angles' of the 
scattering clouds, as shown in Figure~\ref{fig-clump}a.  If we obtain 
the viewing angles of each cloud, we can locate each cloud along the 
line of sight with respect to the central radiation source, and thus 
we will be able to have a three-dimensional view of this nuclear 
region.

We simply assume here that the UV radiation consists of only radiation 
scattered by free electrons and no other diluting component exists.  
This is based on the following two observational arguments in 
NGC~1068.

(i) \hst/FOS UV spectropolarimetry shows that the polarization is 
almost constant over the UV range and its degree declines only in the 
wavelength longer than $\sim 2800$\AA\ (Antonucci et al.  1994).  
This has been interpreted to mean that the starlight contribution is 
significant only in the redward of $2800$\AA, almost outside the F253M 
filter transmission.

(ii) The same FOS spectropolarimetry and also the ground-based 
spectropolarimetry shows no evidence of any enhancement in 
polarization in the broad lines relative to the nuclear continuum.  
The broad lines are believed to originate only in a region much more 
compact than the scattering region that we have seen in our imaging 
polarimetry data.  Any diluting radiation such as starlight or 
free-free emission from the scattering region itself (Tran 1995c), 
only dilutes the polarization of the scattered continuum, not the 
scattered broad lines.  Therefore, if diluting radiation were present, 
the polarization would be higher in the broad lines than in the 
continuum, which has been observed in several Seyfert~2 galaxies, but 
not in NGC~1068 (Antonucci et al.  1994; Tran 1995a,b).

Those spectropolarimetric data, however, cover a large area, $4.''3 
\times 1.''4$ aperture in Antonucci et al.'s data and a $2.''4$ wide 
slit in Tran's data, which could mean that the assumption of 
scattered-light domination might be true only in the brightest clouds.  
Therefore we restrict ourselves to use only the polarization degree of 
the bright knots.  We will discuss the validity of this assumption 
again in \S~\ref{s-disc-3D}.

The following method for obtaining the three-dimensional gas 
distribution and its result have been discussed in Kishimoto (1997).  
We present the method and results here, with the errors in the observed 
polarization that have been estimated in much more detail, and with 
the position of the nucleus that has been obtained more accurately.

\subsection{Method}
\label{s-3D-basis}
The polarization of the scattered radiation from a cloud depends 
basically on its viewing angle, but also on the size of the cloud 
compared to the distance to the illuminating source.  This effect is 
large if the cloud is very close to the source.  In this section, we 
describe a simple method for interpreting the degree of polarization, 
taking this effect into account.

Consider a spherical, uniform cloud of radius $\rc$, which is 
optically thin to scattering.  The cloud is illuminated by a point 
source from a distance $r$ (see Figure~\ref{fig-clump}a).  We consider 
only electron scattering, because the primary scatterers in the 
innermost region of NGC~1068 are thought to be electrons (Antonucci \& 
Miller 1985; Miller, Goodrich \& Matthews 1991).  The calculation of 
polarization for optically thin systems of axisymmetric distribution 
has been implemented by Brown \& McLean (1977).  We follow their 
naming conventions for a geometrical factor $\alpha$ and shape factor 
$\gamma$.  Our three quantities, polarization of the scattered 
radiation $P$, viewing angle $\psi$, and the relative cloud radius 
$\eta \equiv \rc/r$, are related as
\begin{equation}
	P=\frac{-\sin^{2}\psi}{2\alpha(\eta)+\sin^{2}\psi},
	\label{eq-pol}
\end{equation}
where $\alpha(\eta)$ is a function of $\eta$ and defined as
\begin{equation}
	\alpha \equiv (1 + \gamma)/(1 - 3\gamma).
	\label{eq-alpha}
\end{equation}
We find the shape factor $\gamma$ as
\begin{equation}
	\gamma = \frac{\eta^{3}}{2k(\eta)} + \frac{1 - \eta^{2}}{4},
	\label{eq-gamma}
\end{equation}
where 
\begin{equation}
	k(\eta) = \eta - (1 - \eta^{2}) \ln \frac{1+\eta}{\sqrt{1-\eta^{2}}}.
	\label{eq-k}
\end{equation}
Since the projected distance $\rp \equiv r \sin \psi$ is the 
observational quantity whereas the actual distance $r$ is not, we 
re-write $\eta$ as
\begin{equation}
	\eta = \eta' \sin \psi, {\rm \ where \ } \eta' \equiv \rc/\rp.
	\label{eq-eta}
\end{equation}
These relations are rather complicated, but for small~$\eta$, 
$\gamma$ becomes close to unity, hence $\alpha$ close to $-1$, and 
eq.(\ref{eq-pol}) becomes the well-known equation for Thomson 
scattering as

\begin{equation}
	P=\frac{1-\cos^{2}\psi}{1+\cos^{2}\psi}.
	\label{eq-thomson}
\end{equation}
This means that the shape of the cloud or the distribution of the gas 
inside the cloud does not affect the polarization of the scattered 
radiation if the cloud size is small enough compared with the distance 
to the illuminating source, and the polarization depends only on the 
viewing angle, as a natural consequence.

The three-dimensional positions of the clouds can now be obtained, 
given the position of the central source in the image $(x_{c}, 
y_{c})$.  If we denote the position of the center of each cloud in the 
image as $(x,y)$, the distance $z$ for each cloud from the sky plane 
in which the nucleus resides (see Fig.~\ref{fig-clump}b) is written as
\begin{equation}
	z =	\rp /\tan\psi, 
	\label{eq-z}
\end{equation}
where
\begin{equation}
	\rp = \sqrt{ (x-x_{c})^{2} + (y-y_{c})^{2} }.
\end{equation}

One of the ambiguities in this method is that, due to the 
forward-backward symmetry of the optically thin electron scattering, 
there are two possible viewing angles derived from a single value of 
polarization, i.e. the forward and backward scattering case.  These 
two angles correspond to the two positions along the line of sight, 
shown as Figure~\ref{fig-clump}b.  Also, if a cloud is not 
sufficiently resolved, several clumps may exist along one line of 
sight.  This could be the limit of our method, but in this case, the 
two positions derived from the polarization set a constraint on the 
positions of the clouds.  If there is a scattering clump between these 
two positions, there must be another one outside of these two 
positions, to dilute the highly polarized scattered light from the 
inner clump.  In this sense, these two positions should be considered 
as the ``critical'' positions.  Having these ideas in mind, we simply 
assume that there is only one cloud along one line of sight.

\subsection{3D mapping}\label{s-3D-map}
\begin{figure}[bt]
\epsscale{1.0}
\plotone{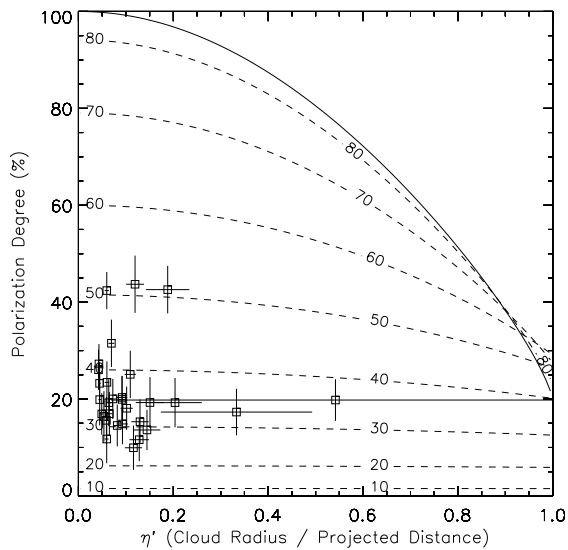}
\figcaption{\label{fig-view} {\capsize
The map of viewing angle $\psi$ on the 
plane of $(\eta', P)$.  The observed positions of each cloud are also 
shown as squares with errors.}}
\end{figure}

Given the position of the nucleus $(x_{c},y_{c})$, the viewing angle 
$\psi$ can be calculated numerically from the two observational 
quantities, polarization $P$ and the relative size of cloud $\eta'$, 
using equations~(\ref{eq-pol})$\sim$(\ref{eq-eta}).  
Figure~\ref{fig-view} shows the map of the viewing angle on the plane 
of $(\eta', P)$ for a forward scattering case.  There is an upper 
limit in $P$ for certain $\eta'$ when $\psi$ is $90\dgr$.  This 
maximum polarization becomes smaller for larger $\eta'$.  This is 
regarded as a ``geometrical dilution'' of polarization, where the 
position angles of polarization of the scattered light from different 
places inside the cloud become significantly different so that the 
polarization tends to cancel out.  As a consequence, there is a 
``region of uncertainty'' at around $\eta' = 0.9 \sim 1.0$ and $P = 25 
\sim 35\%$, where two viewing angles can be solutions for a single 
point of $(\eta', P)$.  For the observational points in this region, 
with certain values for the errors in $\eta'$ and $P$, we can obtain 
the viewing angle only with a large uncertainty.

\begin{figure}[bt]
\epsscale{1.0}
\plotone{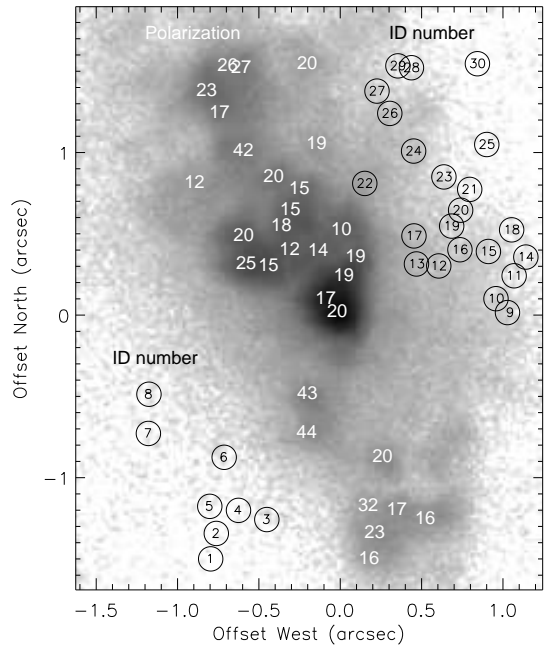}
\figcaption{\label{fig-pol} {\capsize
Observed polarization degree of each cloud 
is shown on the center of each cloud in white numbers.  The ID numbers 
for each cloud are also shown as black numbers in circle, displaced to 
the east and west for lower and upper positions, respectively.}}
\end{figure}

In Figure~\ref{fig-view}, we have plotted the observational points of 
the clumps assuming that the cloud radius is our aperture radius 
$0.''15 /2$, with the total error in $P$ estimated in 
\S~\ref{s-data-err}, and the error in $\eta'$ that originates from the 
error in the position of the nucleus.  We have taken the error of the 
nuclear position symmetrically as $(-0.''03 \pm 0.''05, -0.''12 \pm 
0.''10)$ for simplicity.  Except the central few clouds, we see the 
viewing angles are well approximated by equation~(\ref{eq-thomson}) 
which is the relation for $\eta' = 0$.  This means that the cloud 
shape or gas distribution inside the cloud does not affect the result 
as described above, although we have used equation~(\ref{eq-pol}) to 
derive the viewing angles for all the clumps.  The polarization for 
each cloud is shown in Figure~\ref{fig-pol}.  The results are 
summarized in Table~\ref{table-cloud}, with equatorial coordinates as 
in Table~\ref{table-pos}.

\begin{deluxetable}{crrrrrrrr}
	\footnotesize
	\tablecaption{Polarization and 3D position of cloud
	  \label{table-cloud}}
	\tablewidth{0pt}
	\tablehead{
	\colhead{ID No.}   & \colhead{$x\ ('')$} & 
	\colhead{$y\ ('')$\tablenotemark{a}} & 
	\colhead{R.A. (J2000)} & \colhead{Dec. (J2000)} &
	\colhead{$P$ (\%)} & \colhead{$\theta_{\rm PA} (\dgr)$} &
	\colhead{$\psi\ (\dgr)$} & \colhead{$z\ ('')$}
	}
	\startdata
 1 &  0.19 & -1.50 &  02 42 40.698 & -00 00 49.31 & $16.4 \pm 5.2$ & $ 96.6 \pm\s 8.9$ & $32.1 \pm 4.9$ & $ 2.23 \pm 0.42$ \\
 2 &  0.23 & -1.34 &        40.696 &        49.15 & $23.4 \pm 4.3$ & $104.9 \pm\s 6.6$ & $38.1 \pm 3.4$ & $ 1.60 \pm 0.19$ \\
 3 &  0.54 & -1.26 &        40.675 &        49.07 & $15.6 \pm 4.8$ & $122.3 \pm\s 9.0$ & $31.3 \pm 4.7$ & $ 2.09 \pm 0.38$ \\
 4 &  0.36 & -1.20 &        40.687 &        49.01 & $17.0 \pm 4.5$ & $103.5 \pm\s 8.5$ & $32.6 \pm 4.1$ & $ 1.80 \pm 0.29$ \\
 5 &  0.19 & -1.18 &        40.698 &        48.99 & $31.5 \pm 4.9$ & $ 95.4 \pm\s 5.1$ & $43.8 \pm 3.2$ & $ 1.12 \pm 0.13$ \\
 6 &  0.28 & -0.87 &        40.693 &        48.68 & $20.3 \pm 4.5$ & $106.4 \pm\s 7.4$ & $35.6 \pm 3.7$ & $ 1.14 \pm 0.16$ \\
 7 & -0.19 & -0.73 &        40.724 &        48.54 & $43.7 \pm 5.9$ & $ 71.5 \pm\s 3.5$ & $51.4 \pm 3.4$ & $ 0.50 \pm 0.06$ \\
 8 & -0.19 & -0.49 &        40.723 &        48.30 & $42.6 \pm 4.9$ & $ 62.8 \pm\s 3.8$ & $50.9 \pm 2.9$ & $ 0.32 \pm 0.03$ \\
 9 &  0.00 &  0.02\tablenotemark{b} &        40.711 &        47.79 & $19.8 \pm 4.3$ & $ 63.2 \pm\s 6.3$ & $36.0 \pm 3.9$ & $ 0.19 \pm 0.03$ \\
10 & -0.07 &  0.10 &        40.716 &        47.71 & $17.3 \pm 4.8$ & $ 91.2 \pm\s 7.3$ & $33.2 \pm 4.5$ & $ 0.34 \pm 0.06$ \\
11 &  0.04 &  0.24 &        40.708 &        47.57 & $19.3 \pm 5.1$ & $ 86.4 \pm\s 6.4$ & $34.7 \pm 4.4$ & $ 0.53 \pm 0.09$ \\
12 & -0.42 &  0.30 &        40.739 &        47.51 & $15.3 \pm 4.6$ & $138.4 \pm\s 7.5$ & $31.1 \pm 4.5$ & $ 0.96 \pm 0.17$ \\
13 & -0.56 &  0.31 &        40.748 &        47.50 & $25.1 \pm 4.9$ & $139.2 \pm\s 4.9$ & $39.3 \pm 3.7$ & $ 0.84 \pm 0.11$ \\
14 &  0.11 &  0.36 &        40.704 &        47.45 & $19.3 \pm 5.3$ & $ 74.7 \pm\s 6.3$ & $34.7 \pm 4.5$ & $ 0.72 \pm 0.12$ \\
15 & -0.12 &  0.39 &        40.719 &        47.42 & $13.6 \pm 4.2$ & $ 99.5 \pm\s 9.3$ & $29.3 \pm 4.3$ & $ 0.92 \pm 0.16$ \\
16 & -0.29 &  0.40 &        40.731 &        47.41 & $11.6 \pm 4.4$ & $125.6 \pm  10.0$ & $27.2 \pm 5.0$ & $ 1.14 \pm 0.25$ \\
17 & -0.58 &  0.49 &        40.749 &        47.32 & $19.9 \pm 4.8$ & $135.2 \pm\s 6.1$ & $35.2 \pm 4.1$ & $ 1.16 \pm 0.18$ \\
18 &  0.02 &  0.52 &        40.709 &        47.29 & $ 9.9 \pm 4.6$ & $ 80.3 \pm  11.0$ & $25.2 \pm 5.6$ & $ 1.37 \pm 0.35$ \\
19 & -0.35 &  0.55 &        40.734 &        47.26 & $18.1 \pm 4.5$ & $121.7 \pm\s 7.2$ & $33.6 \pm 4.0$ & $ 1.11 \pm 0.17$ \\
20 & -0.29 &  0.65 &        40.730 &        47.16 & $14.9 \pm 4.2$ & $111.4 \pm\s 9.0$ & $30.6 \pm 4.2$ & $ 1.37 \pm 0.23$ \\
21 & -0.23 &  0.77 &        40.727 &        47.04 & $14.5 \pm 4.3$ & $111.9 \pm\s 9.5$ & $30.3 \pm 4.3$ & $ 1.57 \pm 0.27$ \\
22 & -0.88 &  0.81 &        40.770 &        47.00 & $11.7 \pm 5.0$ & $127.0 \pm  10.9$ & $27.3 \pm 5.6$ & $ 2.44 \pm 0.59$ \\
23 & -0.39 &  0.85 &        40.737 &        46.96 & $20.0 \pm 4.1$ & $105.7 \pm\s 7.1$ & $35.3 \pm 3.5$ & $ 1.46 \pm 0.19$ \\
24 & -0.58 &  1.01 &        40.750 &        46.80 & $42.4 \pm 3.8$ & $112.1 \pm\s 4.1$ & $50.5 \pm 2.2$ & $ 1.03 \pm 0.08$ \\
25 & -0.13 &  1.05 &        40.720 &        46.76 & $19.3 \pm 4.9$ & $ 96.0 \pm\s 7.6$ & $34.6 \pm 4.2$ & $ 1.70 \pm 0.27$ \\
26 & -0.73 &  1.24 &        40.759 &        46.57 & $16.9 \pm 4.3$ & $115.4 \pm\s 8.2$ & $32.6 \pm 4.0$ & $ 2.39 \pm 0.36$ \\
27 & -0.80 &  1.38 &        40.765 &        46.43 & $23.2 \pm 4.2$ & $115.2 \pm\s 6.2$ & $37.9 \pm 3.2$ & $ 2.17 \pm 0.25$ \\
28 & -0.59 &  1.52 &        40.750 &        46.29 & $27.2 \pm 4.1$ & $105.5 \pm\s 5.7$ & $40.9 \pm 2.9$ & $ 2.00 \pm 0.21$ \\
29 & -0.68 &  1.53 &        40.756 &        46.28 & $26.0 \pm 4.1$ & $109.4 \pm\s 6.0$ & $40.0 \pm 3.0$ & $ 2.11 \pm 0.23$ \\
30 & -0.19 &  1.55 &        40.723 &        46.26 & $19.9 \pm 5.3$ & $ 85.1 \pm\s 6.9$ & $35.2 \pm 4.5$ & $ 2.38 \pm 0.40$ \\
	\enddata 
	\tablenotetext{a}{Measured from the UV peak, which is 
	taken as the peak position in the `sharp UV' image (with fine 
	focus).} 
	\tablenotetext{b}{The position of the maximum intensity 
	in the $I$ image (total intensity, out of focus)
	is slightly different from that in the `sharp UV image'.}
\end{deluxetable}

The errors in $\psi$ and $z$ are calculated only from the error in $P$ 
in Table~\ref{table-cloud} and we did not include the error in the 
nucleus position, since it has a systematic effect on {\it all} 
clouds.  The polarization degrees have been debiased following Simmons 
\& Stewart (1985).  That is, the final S/N in polarization, 
$P/\sigma_{P}$, where $\sigma_{P}$ is the total uncertainties in $P$ 
as estimated in \S~\ref{s-data-err}, is large enough for all the 
clumps so that we are allowed to use the equation given in Wardle \& 
Kronberg (1974),
\begin{equation}
	P_{\rm corrected} = P_{\rm obs} 
	  \sqrt{1 - \left( \frac{\sigma_{P}}{P_{\rm obs}} \right)^{2} },
	\label{eq-pcor}
\end{equation}
where $P_{\rm obs}$ is calculated simply as $\sqrt{Q^{2}+U^{2}}/I$.

\begin{figure}[bt]
\epsscale{1.0} 
\plotone{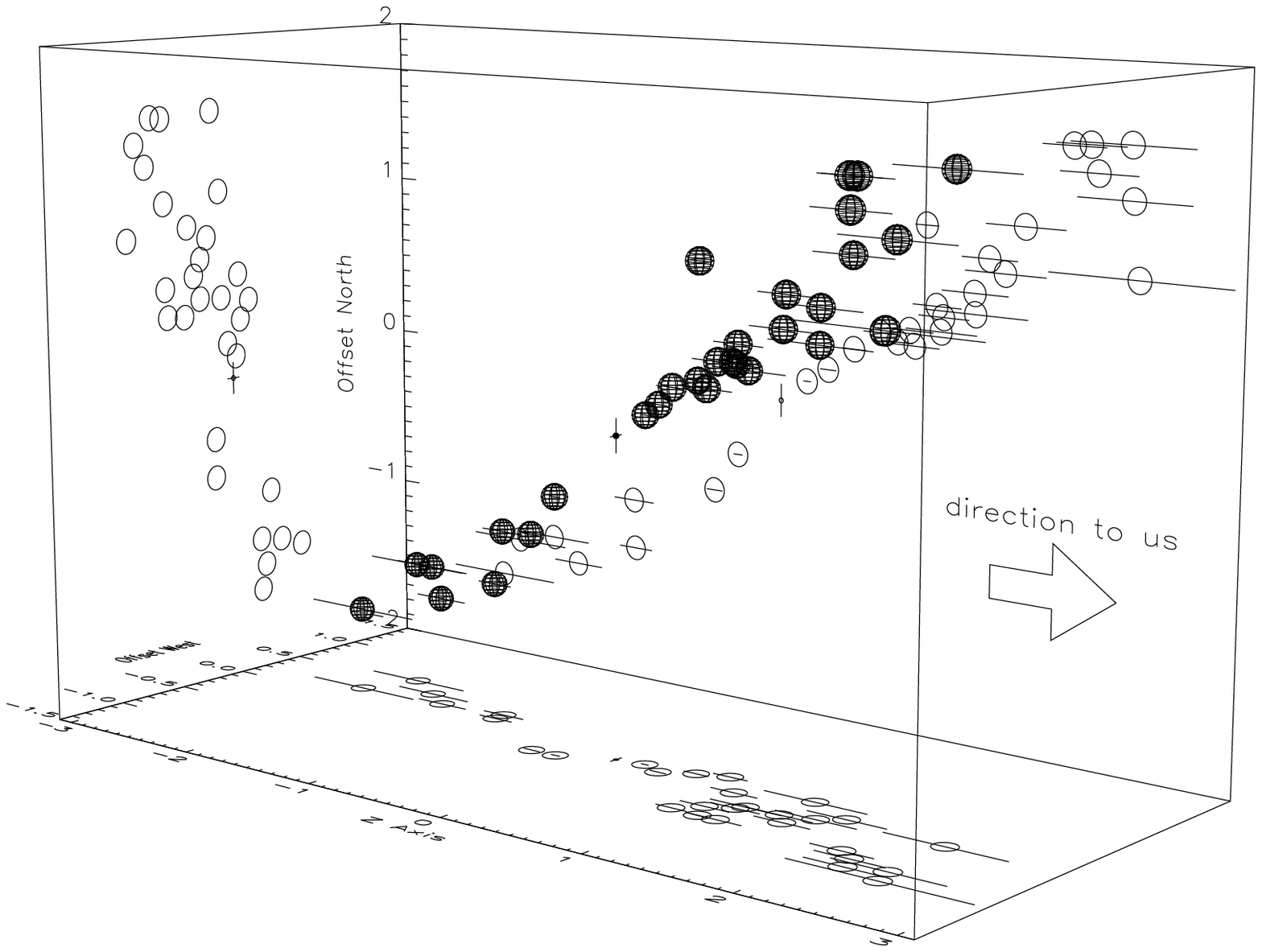} 
\figcaption{\label{fig-3D} 
{\capsize Three-dimensional view of the nuclear region.  Each cloud is 
drawn as a sphere of diameter $0.''15$, and the nucleus is shown as a 
smaller sphere in the center, with the errors in its position.  The 
scale unit is one arcsecond.  The projected views are also illustrated 
on each coordinate plane with each cloud drawn as an open circle.  The 
direction to us is indicated by the large white arrow.  The error in 
the position of the cloud center along the line of sight is indicated 
by the solid line for each cloud.}}
\end{figure}

From these data, we have constructed the three-dimensional view of the 
nuclear region in Figure~\ref{fig-3D}. Each cloud is illustrated 
as a sphere with $0.''15$ diameter.  The position of the nucleus is 
shown as a smaller sphere with its error.  The solid line at each 
cloud indicates the error of its position from the uncertainties in 
$P$.  The positions in this figure would be only for one of the two
possible cases.  We could select another side of the critical 
positions for each cloud, or even both.  However, as we also discuss in 
the next section, there is observational evidence that suggest an 
absorption excess in the southern region compared to the northern 
region, from a color difference (Macchetto et al.  1994) and HI 
absorption (Gallimore et al.  1994).  Therefore we have simply 
selected the case shown in Figure~\ref{fig-3D}.

\section{Discussion}\label{s-disc}
\subsection{PA map}\label{s-disc-PA}
We have shown that the UV polarization image is completely consistent 
with a point-source scattering within the accuracy of the FOC 
polarimetry, over wide regions from $\sim$100 pc scale down to 
$\sim$10 pc scale around the nucleus.  The scattered radiation is very 
extended to the east and west, as well as to the north and south, as 
shown in Figure~\ref{fig-PA-B}.  We should note, however, that there 
could be much smaller deviations from a point-source scattering, which 
we cannot discuss with the given limit of the accuracy of these 
imaging polarimetry data.  We just certainly do not see any 
significant deviations from a centrosymmetric pattern beyond the 
estimated errors of the data, when we exclude the regions contaminated 
by the effect of the bad focus.

\begin{figure}[bt]
\epsscale{1.0}
\plotone{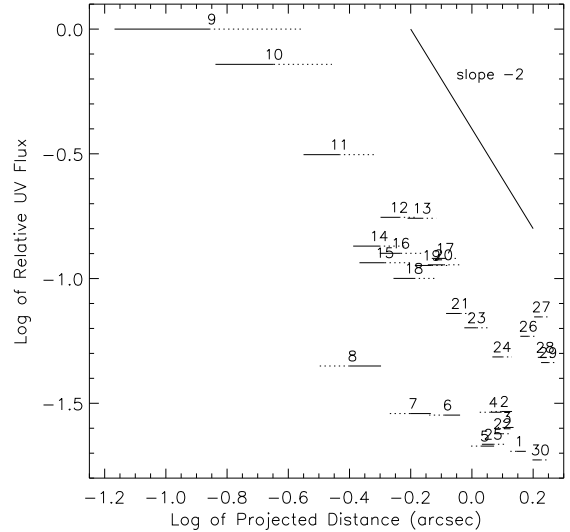}
\figcaption{\label{fig-flux} {\capsize
The relative UV flux for each cloud in 
Fig.~\ref{fig-pol} is plotted as its ID number, against the projected 
distance from the nucleus, in log scale.  The horizontal lines 
represent the error in the projected distance, where the solid 
(dotted) ones are for the case with the nucleus shifted to the north 
(south).  The slope of $-2$ is indicated in the upper-right.}}
\end{figure}

Young et al.  (1996) have estimated the size of the `torus' that is 
thought to be obscuring the nucleus.  They have calculated its size to 
be greater than 200 pc ($\sim 3''$) in diameter, based on the 
absorption feature at $\sim 1''$ south from the nucleus in their 
near-infrared polarized flux image.  In Figure~\ref{fig-flux}, we have 
plotted the UV intensities of each cloud against the projected 
distance of each cloud from the nucleus, with the error of the 
projected distance originating from the error in the position of the 
nucleus.  The statistical errors of the UV intensities were quite 
small, so are not shown in the figure.  We clearly see that the fluxes 
of the southern clouds (Nos.1$\sim$8) are systematically smaller than 
those of the northern clouds, which suggests an absorption excess in 
the south.  However, if we assume that the radial distribution of the 
UV flux is the same in the south as in the north, the observed UV flux 
ratio suggests an absorption of only $A_{V} \lesssim 1$ at $\sim 0.''5 
- 1''$ south from the nucleus.  The obscuring `torus' could be 
extended to this scale, but it should be much smaller if we define it 
as the absorbing material which has enough optical thickness to hide 
the scattered radiation completely.  The size of this material must be 
less than $0.''3$ projected onto the sky which corresponds to $\sim$ 
20 pc, based on the distribution of the polarization seen in 
Figure~\ref{fig-PA-A}.

\begin{figure}[bt]
\epsscale{1.0} 
\plotone{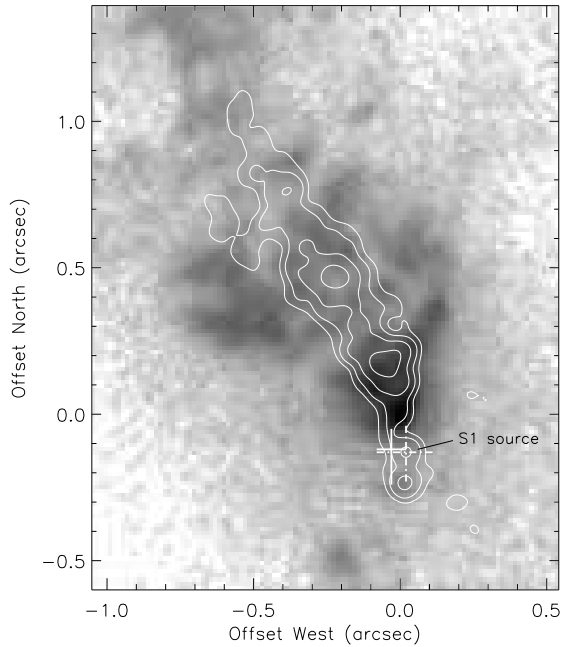} 

\figcaption{\label{fig-radio} {\capsize The registration of the 5-GHz 
MERLIN radio map (Muxlow et al.  1996) onto the \hst\ sharp UV image, 
using the astrometric result of Capetti et al.  (1997).  Both are in 
log scale.  The registration uncertainty is shown as a dash-dot plus 
at the S1 source.  Our new location of the nucleus is drawn as a 
solid plus, whose size represents our estimated error. }}

\end{figure}

Muxlow et al.  (1996) and Gallimore et al.(1996c) argue that the most 
probable location of the nucleus in the VLBI radio image is one of the 
most southern components, called the S1 source (Gallimore et al.  
1996a).  Strong water maser sources have been found at this S1 source 
(Gallimore et al.  1996b; Greenhill et al.  1996; Greenhill \& Gwinn 
1997), and these sources have been interpreted to be associated with 
the obscuring torus around the nucleus.  In Table~\ref{table-pos}, we 
show the offset of the S1 source from the UV peak, taken from the 
absolute astrometric result of Capetti et al.  (1997).  In 
Figure~\ref{fig-radio}, the 5-GHz MERLIN radio map by Muxlow et 
al.(1996) is registered on the sharp UV image (F253M filter, fine 
focus), using this astrometric result.  The uncertainty of the 
registration is indicated at the S1 source.  Our new location of the 
nucleus is shown with the error stated in Table~\ref{table-pos}.  The 
new location coincides with the S1 source within the errors.  
Therefore our result supports that the hidden nucleus resides in the 
S1 source.

\subsection{3D distribution}\label{s-disc-3D}
The largest source of ambiguity in the derived 3D distribution is the 
assumption of scattered-light domination in the UV range.  Both of 
the two observational results described in \S~\ref{s-3D-assume} are 
for large apertures and our assumption may not be valid especially in 
the faint regions.

We can roughly check this scattered-light domination by looking at the 
radial dependence of the UV flux in Figure~\ref{fig-flux}.  If we 
assume that the gas density, volume, and viewing angle are the same 
for all the clumps, the plotted points should distribute with a slope 
of $-2$.  The overall distribution mostly follows this slope.  
However, some of the clumps are above the line. This could be 
compensated by the differences in density and volume, but it also 
would suggest the existence of diluting radiation (non-scattered, 
unpolarized light) in these clouds.

The determination of the viewing angle of the UV brightest cloud (UV 
peak) could be beyond the limit of our method, since the cloud is too 
close to the nucleus.  In this case, inhomogeneity of the cloud within 
the aperture will have a significant effect on the polarization, while 
these imaging polarimetry data do not have enough spatial resolution 
for further investigation.  This uncertainty, however, does not affect 
the overall 3D distribution in Figure~\ref{fig-3D}.

A correlation between the optical morphology in the \hst\ images and 
the radio jet structure (Muxlow et al.  1996; Gallimore et al.  
1996a,c; see Fig.\ref{fig-radio}) has been suggested by Capetti et al.  
(1997).  Our 3D distribution obtained might indicate a certain linear 
structure, which could be related to the jet structure.  However, much 
more detailed consideration of the nature of the UV radiation is 
needed.  Detailed analysis using other \hst\ images, including the 
discussion of the fraction of diluting radiation, will be presented in 
a subsequent paper (Kishimoto 1998).

\section{Conclusions}\label{s-conc}
We have analyzed the \hst/FOC imaging polarimetry data of NGC~1068, 
and shown that the data are consistent with a simple point-source 
scattering model, within the FOC polarimetric accuracy.  We have 
re-determined the location of the nucleus, by eliminating the regions 
which are suspected to have significant contaminations from the 
neighboring pixels.  The error circle suggests that the nucleus is 
located between clouds~A and~B, and the most probable location of the 
nucleus has been found to be very close to cloud B, as close as $\sim 
0.''08$, which is much closer than has ever been claimed before.

Based on this result, we have derived the three-dimensional view of 
the nuclear gas distribution, assuming that the UV radiation is 
dominated by scattered radiation in the bright knots.  The inferred 
three-dimensional distribution of the clouds might suggest the 
existence of a linear structure which could be related to the radio 
jet, though the assumption of scattered-light domination should be 
examined using other high-resolution images.

\acknowledgments The author thanks Robert Jedrzejewski for providing 
various information on the FOC polarimetry, and Robert Antonucci for 
his helpful suggestions to improve the manuscript.  The author would 
like to thank Ryuko Hirata for his kind advice and generous support.  
The author also appreciates discussions with David Axon and Alessandro 
Capetti, and would like to thank Jack Gallimore for kindly providing 
the electronic data for the MERLIN radio map.

\appendix
\section{Error Estimation}
In this appendix we summarize the method for the error estimation 
discussed in \S~\ref{s-data-err}.  

Consider the incident radiation of the Stokes parameters $(I,Q,U,V)$.  
The flux through three polarizers, $f_{i} (i = 1,2,3)$, are written as
\begin{equation}
	f_{i} = \frac{1}{2} t_{i} 
	  (I - k_{i}\cos 2\theta_{i}\cdot Q 
	     - k_{i}\sin 2\theta_{i}\cdot U),
	\label{eq-fi}
\end{equation}
where $t_{i}, k_{i}$, and $\theta_{i}$ are each polarizer's 
transmittance, polarization efficiency, and axis direction, 
respectively.  We define the polarizers' axis directions to be 
measured counter-clockwise from the $+x$ axis direction of the image, 
while we set the reference plane of the Stokes parameter along the $y$ 
axis so that the position angle of the polarization is measured 
counter-clockwise from the $y$ direction.

The Stokes parameters $(I,Q,U)$ are derived from the observed 
fluxes $f_{i}$ through each polarizer by the inverse relation of 
the above equation.  For convenience, we define
\begin{equation}
    f_{i}' \equiv \frac{2}{t_{i}} f_{i},
    \label{eq-fid}
\end{equation}
which corresponds to the transmittance-corrected incident flux for 
unpolarized light.  If we write $(I,Q,U)$ also as 
$(I_{1},I_{2},I_{3})$, these are calculated as
\begin{equation}
	I_{i} = \sum_{j=1}^{3}a_{ij}f_{j}',
	\label{eq-stokes}
\end{equation}
where
% \begin{equation}
% 	[a_{ij}] = \frac{1}{A} \cdot \\
% 	\left[
% 	  \begin{array}{ccc}  
% 	     k_{2}k_{3}\sin(-2\theta_{2}+2\theta_{3}) & 
% 	     k_{3}k_{1}\sin(-2\theta_{3}+2\theta_{1}) &
% 	     k_{1}k_{2}\sin(-2\theta_{1}+2\theta_{2}) \\
% 	    -k_{2}\sin2\theta_{2}+k_{3}\sin2\theta_{3} &
% 	    -k_{3}\sin2\theta_{3}+k_{1}\sin2\theta_{1} &
% 	    -k_{1}\sin2\theta_{1}+k_{2}\sin2\theta_{2} \\
% 	     k_{2}\cos2\theta_{2}-k_{3}\cos2\theta_{3} &
% 	     k_{3}\cos2\theta_{3}-k_{1}\cos2\theta_{1} &
% 	     k_{1}\cos2\theta_{1}-k_{2}\cos2\theta_{2} 
% 	  \end{array}
% 	\right]
% 	\label{eq-aij}
% \end{equation}
\begin{eqnarray}
	\left[ 
	  \begin{array}{c}
	    a_{11} \\ a_{21} \\ a_{31}
	  \end{array}
	\right]  & = & \frac{1}{A}
	\left[
	  \begin{array}{l}  
	   {\?}k_{2}k_{3}\sin(-2\theta_{2}+2\theta_{3})  \\
	   -   k_{2}\sin2\theta_{2}+k_{3}\sin2\theta_{3} \\
	   {\?}k_{2}\cos2\theta_{2}-k_{3}\cos2\theta_{3}
	  \end{array}
	\right] \nonumber\\ 
	\medskip
	\left[ 
	  \begin{array}{c}
	    a_{12} \\ a_{22} \\ a_{32}
	  \end{array}
	\right]  & = & \frac{1}{A}
 	\left[
 	  \begin{array}{l}  
       {\?}k_{3}k_{1}\sin(-2\theta_{3}+2\theta_{1})  \\
 	   -   k_{3}\sin2\theta_{3}+k_{1}\sin2\theta_{1} \\
 	   {\?}k_{3}\cos2\theta_{3}-k_{1}\cos2\theta_{1}
 	  \end{array}
 	\right] \nonumber\\
	\medskip
	\left[ 
	  \begin{array}{c}
	    a_{13} \\ a_{23} \\ a_{33}
	  \end{array}
	\right]  & = & \frac{1}{A}
 	\left[
 	  \begin{array}{l}  
 	    {\?}k_{1}k_{2}\sin(-2\theta_{1}+2\theta_{2})  \\
 	    -   k_{1}\sin2\theta_{1}+k_{2}\sin2\theta_{2} \\
 	    {\?}k_{1}\cos2\theta_{1}-k_{2}\cos2\theta_{2}
 	  \end{array}
 	\right]
 	\label{eq-aij}
\end{eqnarray}
and 
\begin{eqnarray}
	A & = &    k_{2}k_{3}\sin(-2\theta_{2}+2\theta_{3}) \nonumber\\
	 && {\ } + k_{3}k_{1}\sin(-2\theta_{3}+2\theta_{1}) \nonumber\\ 
	 && {\ } + k_{1}k_{2}\sin(-2\theta_{1}+2\theta_{2}).
	\label{eq-A}
\end{eqnarray}
Note that for the nominal case of $\theta_{1}=180\dgr, 
\theta_{2}=60\dgr, \theta_{3}=120\dgr$, and $k_{1}=k_{2}=k_{3}=1$, 
this equation simply becomes
\begin{equation}
	\left[ 
	  \begin{array}{c}
	    I_{1} \\ I_{2} \\ I_{3}
	  \end{array}
	\right]	   =  \frac{1}{3}
 	\left[
 	  \begin{array}{lcc}  
 	    {\?}1{\ }  &  1        &  1       \\
 	    -   2{\ }  &  1        &  1       \\
 	    {\?}0{\ }  & -\sqrt{3} & \sqrt{3}
 	  \end{array}
 	\right]
    \left[ 
	  \begin{array}{c}
	    f_{1}' \\ f_{2}' \\ f_{3}'
	  \end{array}
	\right].
 	\label{eq-stokes-simple}
\end{equation}
For the FOC polarizers at the transmission range of F253M filter, we 
take the polarization efficiencies to be (Nota et al. 1996, p38),
\begin{equation}
    k_{1}=0.986, k_{2}=0.976, k_{3}=0.973.
\end{equation}
The polarization degree $P$ and position angle $\theta_{\rm PA}$ are 
calculated from $I, Q, U$ as
\begin{eqnarray}
    P & = & \frac{\sqrt{Q^{2} + U^{2}}}{I}, \nonumber\\
    \theta_{\rm PA} & = & \frac{1}{2} \arctan\frac{U}{Q}.
    \label{eq-PPA}
\end{eqnarray}

Now we estimate the total error in $P$ with the following four major 
sources;
\begin{equation}
    \sigma_{P}^{2} = (\sigma_{P}^{\rm stat})^{2}
               + (\sigma_{P}^{\rm shift})^{2}
               + (\sigma_{P}^{\rm axis})^{2}
               + (\sigma_{P}^{\rm corr})^{2}.
    \label{eq-sig-p}
\end{equation}
The statistical error in the photon counts $\sigma_{P}^{\rm stat}$ is 
a random error, while the other three errors, each described below, 
would be systematic errors.  We treat these three, however, in the 
same way as for random errors, since we do not have enough calibration 
results to correct for them exactly.

The error from the image registration uncertainties, $\sigma_{P}^{\rm 
shift}$, is estimated by actually shifting the images with the 
calibration uncertainties of 0.3 pixel (Hodge 1995) and calculating 
the resulting change of $P$.  The other errors are calculated using 
equations (\ref{eq-fid}) $\sim$ (\ref{eq-PPA}) in the following way.

The statistical error $\sigma_{P}^{\rm stat}$ is written as
\begin{equation}
    (\sigma_{P}^{\rm stat})^{2} = \sum_{i=1}^{3}\left( 
      \frac{\partial P}{\partial f_{i}} \right) ^{2}\sigma_{f_{i}}^{2},
    \label{eq-sig-stat}
\end{equation}
where $\sigma_{f_{i}}$ is calculated assuming poisson noise in the 
counts.  The error from the uncertainties in the directions of 
polarizers' axes, $\sigma_{P}^{\rm axis}$, is calculated as
\begin{equation}
	(\sigma_{P}^{\rm axis})^{2} = \sum_{i=1}^{3}\left( 
      \frac{\partial P}{\partial \theta_{i}} 
      \right) ^{2}\sigma_{\theta_{i}}^{2},
    \label{eq-sig-axis}
\end{equation}
where we take $\sigma_{\theta_{i}}$ to be $3\dgr$ (Nota et.  al 1996, 
p36; see also Robinson \& Thomson 1995).  Finally, $\sigma_{P}^{\rm 
corr}$ is the error from the uncertainties in the ``correction 
factors'', discussed in \S~\ref{s-data-err}, to be multiplied to each 
of $f_{i}$.  To take this into account, we re-write $f_{i}'$ to define 
correction factors $\xi_{i}$ as
\begin{equation}
    f_{i}' = \xi_{i} \cdot \frac{2}{t_{i}} f_{i}
	\label{eq-xi}
\end{equation}
and derive $\sigma_{P}^{\rm corr}$ from
\begin{equation}
	(\sigma_{P}^{\rm corr})^{2} = \sum_{i=1}^{3}\left( 
      \frac{\partial P}{\partial \xi_{i}} 
      \right) ^{2}\sigma_{\xi_{i}}^{2},
    \label{eq-sig-corr}
\end{equation}
where we take $\sigma_{\xi_{i}}$ at most to be 0.05 for the cases (A) 
and (C), 0.04 for the case (B), estimating from three kinds of 
uncertainties discussed in \S~\ref{s-data-err}.  All partial 
derivatives can be calculated from 
equations~(\ref{eq-fid})$\sim$(\ref{eq-PPA}) and (\ref{eq-xi}), 
setting $\theta_{1}=180\dgr, \theta_{2}=60\dgr, \theta_{3}=120\dgr$ 
and $\xi_{i} = 1$.  

The error in $\theta_{\rm PA}$ can also be estimated in just the same 
manner as $\sigma_{P}$ above.  The well-known approximate relation in 
high S/N case,
\begin{equation}
    \sigma_{\theta_{\rm PA}} 
      \simeq \frac{1}{2}\frac{\sigma_{P}}{P} \qquad 
      \mbox{(in radian)},
\end{equation}
holds only for $\sigma_{P}^{\rm stat}$ and $\sigma_{P}^{\rm corr}$ 
(note the similarity between 
$\partial P / \partial f_{i}$ and $\partial P / \partial \xi_{i}$), 
but not for $\sigma_{P}^{\rm shift}$ nor $\sigma_{P}^{\rm axis}$.  In 
particular, $\sigma_{\theta_{\rm PA}}^{\rm axis}$ simply becomes
\begin{equation}
    \sigma_{\theta_{\rm PA}}^{\rm axis} 
      = \frac{1}{\sqrt{2}} \sigma_{\theta}
\end{equation}
when we set the uncertainty in all the three polarizers' axes to be 
$\sigma_{\theta}$ and all three polarization efficiencies to be the 
same.

%%%%% 1/3  REFERENCES %%%%%

\end{document}